\documentclass[aps,prapplied,showkeys,floatfix,reprint]{revtex4-2}

\usepackage{graphicx}
\usepackage{color}
\usepackage{amsmath,amssymb}
\usepackage{bm}
\usepackage{upgreek}
\usepackage{hyperref}
\hypersetup
{
	colorlinks=true,
	allcolors=blue,
}

\newcommand{\x}[1] { \text{#1} }
\newcommand{\e}[1] { \x{e}^{#1} }

\newcommand{\vpte} { V_\x{PTE} }
\newcommand{\ipte} { I_\x{PTE} }
\newcommand{\tel} { T_\x{el} }
\newcommand{\tav} { T_\x{av} }
\newcommand{\cel} { C_\x{el} }
\newcommand{\pin} { P_\x{in} }
\newcommand{\wch} { W_\x{ch} }
\newcommand{\lch} { L_\x{ch} }
\newcommand{\wg} { W_\x{g} }
\newcommand{\tc} { T_\x{c} }

\newcommand{\lc} { L_\x{c} }
\newcommand{\lo} { L_0 }
\newcommand{\leff} { L_\x{eff} }
\newcommand{\lwf} { \mathcal{L} }
\newcommand{\la} { L_\alpha }
\newcommand{\ka} { \kappa }
\newcommand{\ko} { \kappa_0 }
\newcommand{\ef} { E_\x{F} }
\newcommand{\efo} { E_\x{F0} }
\newcommand{\kb} { k_\x{B} }
\newcommand{\vf} { v_\x{F} }

\newcommand{\smin} { \sigma_\x{min} }
\newcommand{\rvopt} { R_V^\x{opt} }

\begin{document}

\title{Optimizing the photothermoelectric effect in graphene}

\author{Aleandro Antidormi}
\email{aleandro.antidormi@icn2.cat}
\affiliation{Catalan Institute of Nanoscience and Nanotechnology (ICN2), CSIC and BIST, Campus UAB, Bellaterra, 08193 Barcelona, Spain}

\author{Aron W. Cummings}
\email{aron.cummings@icn2.cat}
\affiliation{Catalan Institute of Nanoscience and Nanotechnology (ICN2), CSIC and BIST, Campus UAB, Bellaterra, 08193 Barcelona, Spain}

\date{\today}

\begin{abstract}
\noindent Among its many uses, graphene shows significant promise for optical and optoelectronic applications. In particular, devices based on the photothermoelectric effect (PTE) in graphene can offer a strong and fast photoresponse with high signal-to-noise ratio while consuming minimal power. In this work we discuss how to optimize the performance of graphene PTE photodetectors by tuning the light confinement, device geometry, and material quality. This study should prove useful for the design of devices using the PTE in graphene, with applications including optical sensing, data communications, multi-gas sensing, and others.
\end{abstract}

\maketitle

\section{Introduction}

Owing to its unique properties, graphene is a promising material for a wide range of applications \cite{Ferrari2015}. Particularly promising are applications that utilize several of graphene's unique properties in a single system or device. A perfect example of such devices are photodetectors based on the photothermoelectric effect (PTE) in graphene \cite{Lemme2011, Song2011}, which combine a broadband and fast photoresponse with a high signal-to-noise ratio and minimal power consumption \cite{Cai2014, Schuler2016, Koppens2019, Muench2019}.

\begin{figure}[tbh]
\includegraphics[width=\columnwidth]{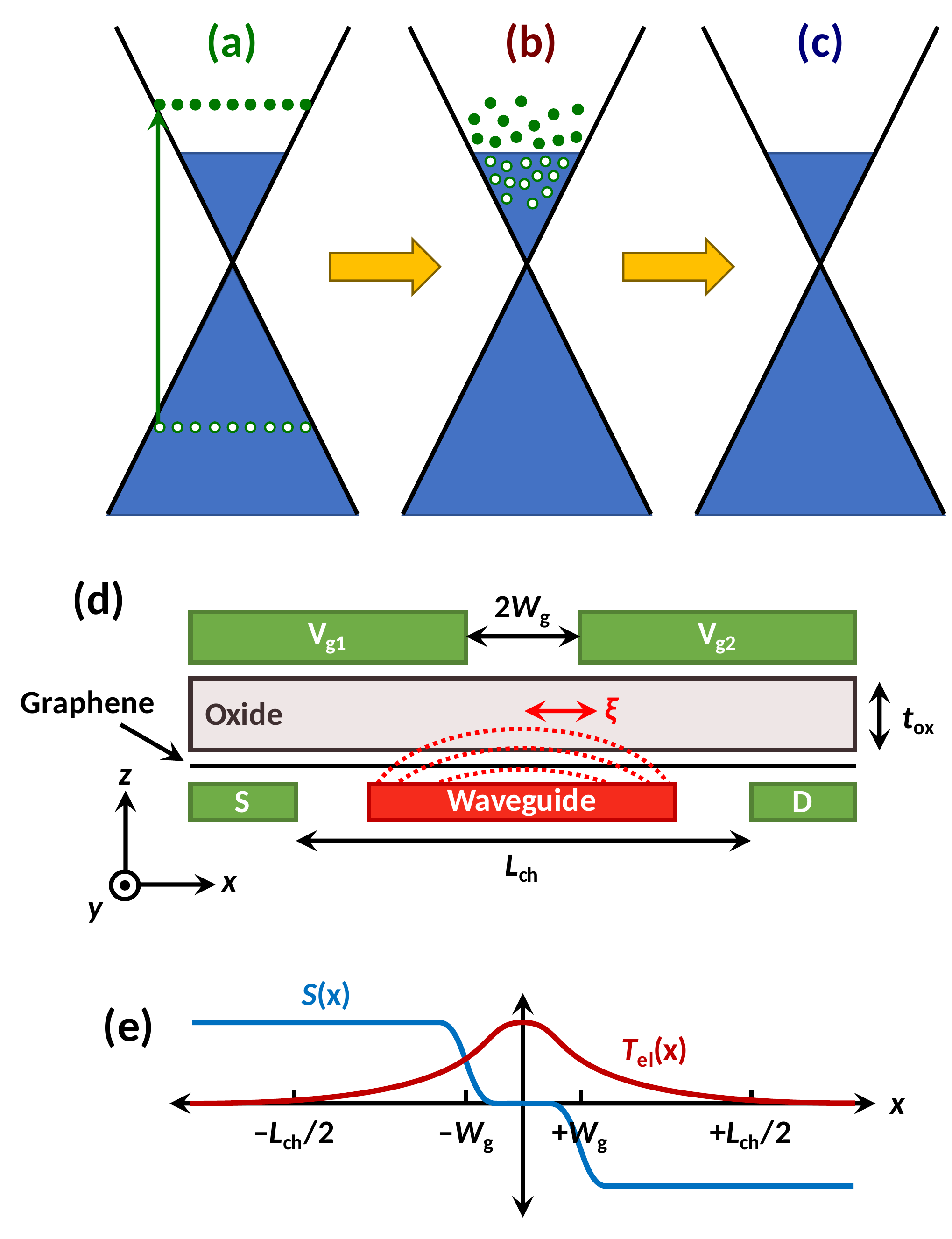}
\caption{Schematic of the photothermoelectric effect in graphene. (a) Optical absorption generates an excited population of electrons and holes, (b) interparticle scattering thermalizes photoexcited carriers to an elevated temperature, and (c) electron-phonon scattering relaxes the electron temperature back to the lattice temperature. Panel (d) shows a cross section of a graphene PTE detector, where split gates generate a \textit{pn} junction in the graphene channel, resulting in a thermoelectric voltage between the source and drain contacts. Panel (e) shows examples of the temperature profile $\tel(x)$ and the Seebeck coefficient $S(x)$ present in these devices.}
\label{fig_pte}
\end{figure}

Figure \ref{fig_pte} illustrates the PTE in graphene and how it may be used for photodetection. Optical absorption first generates a population of excited electrons and holes, as shown in panel (a). Because of its linear electronic band structure \cite{Nair2008}, graphene is a broadband absorber with uniform light absorption from the optical to the THz regime \cite{Dawlaty2008}. Next, interparticle scattering quickly thermalizes these photoexcited carriers to an elevated temperature, as shown in panel (b). Graphene's small electronic heat capacity means this electronic temperature can be much higher than the lattice temperature \cite{Graham2013}, and graphene's fast electronic response means this thermalization process happens within a few tens of femtoseconds \cite{Tielrooij2015}, enabling photodetectors that can operate at very high frequencies \cite{Schuler2016}. Finally, electron-phonon scattering relaxes the electron temperature back to the lattice temperature, as shown in panel (c). The weak interaction between electrons and acoustic phonons in graphene results in a slow decay of the electron temperature \cite{Bistritzer2009}. Defects, disorder, and optical phonon emission can speed up this process \cite{Song2012, Tikhonov2018, Kong2018, Pogna2021}, but even so, the electronic temperature in graphene can remain elevated for several picoseconds, long enough to be efficiently detected \cite{George2008, Lemme2011, Graham2013, Betz2013, Plotzing2014, Laitinen2014, Sierra2015, Halbertal2017, Tomadin2018, Pogna2021}.

This approach to optical detection has been used to design, for example, room-temperature THz detectors with high bandwidth, fast response, and a high signal-to-noise ratio (noise-equivalent power $<$ 40 pW/$\sqrt{\x{Hz}}$) \cite{Cai2014, Koppens2019}; waveguide-integrated photodetectors operating at telecommunications wavelengths that can reach switching frequencies up to and beyond 65 GHz \cite{Schuler2016, Miseikis2020, Marconi2021}; and plasmonically-enhanced detectors reaching photoresponses over 12 V/W that can also be cascaded in series \cite{Muench2019}. Photodetectors based on the PTE in graphene are thus highly promising for their use in many applications. However, there are a large number of design parameters impacting the performance of these devices, including the light profile, device geometry, and material quality.

In this paper, we discuss the impact that these design parameters have on the performance of PTE-based graphene photodetectors, and we demonstrate how their performance may be optimized. We find that by varying the device geometry and material quality within an achievable range, the performance of these photodetectors can be tuned by more than one order of magnitude. These results, and our study of how the performance scales with various geometrical and material parameters, should be highly valuable for the design of future photodetectors based on the PTE in graphene.

\section{Device operation}

Figure \ref{fig_pte}(d) shows a schematic of a PTE-based photodetector. The graphene channel is oriented along the $x$-axis and is situated between source (S) and drain (D) contacts. Light is delivered to the graphene via an optical waveguide oriented along the $y$-axis (out of the page). Absorption of light traversing the width of the graphene channel then leads to a nonuniform electronic temperature profile centered in the middle of the channel (at $x=0$). An example of such a temperature profile is shown in Fig.\ \ref{fig_pte}(e).

This situation differs from typical thermoelectric devices in that the hot ``side'' is the center of the channel while the cold sides are at the source/drain contacts. To generate a nonzero voltage from this temperature profile, it is thus necessary to have a nonuniform Seebeck coefficient along the length of the channel. This is accomplished via the split gates labeled ``$V_\x{g1}$'' and ``$V_\x{g2}$'', which are used to form a \textit{pn} junction that results in a Seebeck coefficient that is antisymmetric across the length of the channel. An example of this nonuniform Seebeck coefficient is shown in Fig.\ \ref{fig_pte}(e).

Finally, this combination of a nonuniform electron temperature and Seebeck coefficient generates a thermoelectric voltage between the source and drain electrodes \cite{Lemme2011, Song2011},
\begin{equation}
\vpte = \int \! S(x) \frac{\x{d}\tel(x)}{\x{d}x} \, \x{d}x,
\label{eq_vpte}
\end{equation}
where $S(x)$ is the spatially dependent Seebeck coefficient and $\tel(x)$ is the electronic temperature relative to the lattice temperature. This voltage may be measured directly in an open-circuit configuration, or measured as a photocurrent in a closed circuit,
\begin{equation}
\ipte = \vpte / R,
\label{eq_ipte}
\end{equation}
where $R$ is the channel resistance.

The signal-to-noise ratio (SNR) of the photodetector may be quantified via the figure of merit known as the noise-equivalent power (NEP), which specifies the input optical power at which the SNR is equal to 1 at a sampling bandwidth of 1 Hz. The PTE-based photodetectors operate at zero bias, such that the only source of noise is Johnson (thermal) noise, and the NEP is thus given by
\begin{equation}
\x{NEP} = \frac{\sqrt{4\kb T R}}{\vpte / \pin}, \label{eq_nep}
\end{equation}
where $\pin$ is the incident optical power, $T$ is the ambient temperature, and $\kb T$ is the thermal energy.

\section{Device optimization}

In this paper we focus on the optimization of three figures of merit: the photovoltage $\vpte$, the photocurrent $\ipte$, and the noise-equivalent power NEP, as defined in Eqs.\ \eqref{eq_vpte}-\eqref{eq_nep}. As seen from these equations, optimization of the photodetectors involves tuning the interplay between the electron temperature profile $\tel(x)$, the Seebeck profile $S(x)$, and the channel resistance $R$.

Our starting point will be the two-dimensional (2D) heat equation, solved within the graphene sheet:
\begin{gather}
\vec{\nabla} \cdot \left[ \ka(x,y) \nabla\tel(x,y) \right] \nonumber \\
- \gamma(x,y) \cel(x,y) \tel(x,y) \nonumber \\
+ P(x,y) = 0,
\label{eq_heat_2d}
\end{gather}
where the $x$-axis ($y$-axis) is parallel (perpendicular) to the graphene channel, $\ka(x,y)$ is the electronic thermal conductivity, $\gamma(x,y)$ is the electron cooling rate, $\cel(x,y)$ is the electronic heat capacity, and $P(x,y)$ is the optical power density incident on the graphene sheet. We assume that $\ka$, $\gamma$, and $\cel$ are uniform along the $y$-axis, and their variation along the $x$-axis is determined by the carrier density induced by the split gates and the native doping of the graphene.

We decompose the optical power density into components along the $x$ and $y$ axes as
\begin{equation}
P(x,y) = \pin \cdot G(x) \cdot \frac{1}{\la}\e{-y/\la},
\end{equation}
where $\pin$ is the total optical power incident on the graphene sheet, $G(x)$ is the profile of the optical power along the length of the channel, and $\la$ is the length over which light is absorbed by the graphene as it traverses the width of the channel, which is determined by the imaginary part of the graphene index of refraction and the degree of coupling to the waveguide, and is typically on the order of 50-100 $\upmu$m \cite{Schuler2016}. In the following sections, we solve the heat equation in different situations to determine the optimal device parameters.

\subsection{Light profile}
\label{sec_light}

In the design of photodetectors, it is known that confining the incoming light can lead to enhanced photoresponse, and this can be achieved in different ways. In one recent waveguide-integrated photodetector, careful design of the waveguide was used to confine a 1.55-$\upmu$m optical mode to a full width of $\sim$80 nm \cite{Schuler2016}. In another approach, the metal split gates above the graphene channel were used to confine the 1.55-$\upmu$m waveguide mode to a full width of $\sim$100 nm \cite{Muench2019}.

Here we quantify the gain in detector performance that arises from confining the incident light profile along the $x$-axis. To isolate this effect, we assume a perfect \textit{pn} junction at $x=0$, such that $S(x) = +S$ $(-S)$ for $x<0$ $(x>0)$. The photovoltage in Eq.\ \eqref{eq_vpte} then reduces to
\begin{equation}
\vpte = 2S \cdot \tel(x=0).
\label{eq_vpte_ideal}
\end{equation}
Here we have assumed that the channel is sufficiently long such that $\tel = 0$ at the source and drain contacts. To find the electron temperature at $x=0$, we start with the 2D heat equation of Eq.\ \eqref{eq_heat_2d}. Because here we are only concerned with transport along the $x$-axis, we reduce to a one-dimensional (1D) heat equation by averaging along the $y$-axis, giving
\begin{gather}
\frac{\x{d}^2\tel(x)}{\x{d}x^2} - \frac{\tel(x)}{\lc^2} + \frac{\alpha \pin}{\ka \wch} G(x) = 0, \nonumber \\
\x{where} \label{eq_heat_x} \\
\alpha = 1 - \exp(-\wch/\la), \nonumber \\
\lc = \sqrt{\ka / \gamma \cel}, \nonumber
\end{gather}
with $\alpha$ the fraction of light absorbed, $\wch$ the channel width, and $\lc$ the electronic cooling length. The cooling length depends on the dominant energy relaxation mechanism in graphene \cite{Song2012, Tikhonov2018, Kong2018, Pogna2021}, but is typically measured to be on the order a few hundred nanometers up to 1 $\upmu$m \cite{Betz2013, Laitinen2014, Sierra2015}. Here, $\lc$ is independent of $x$, owing to the equal and opposite doping of the graphene channel by the split gates.

\begin{figure}[tbh]
\includegraphics{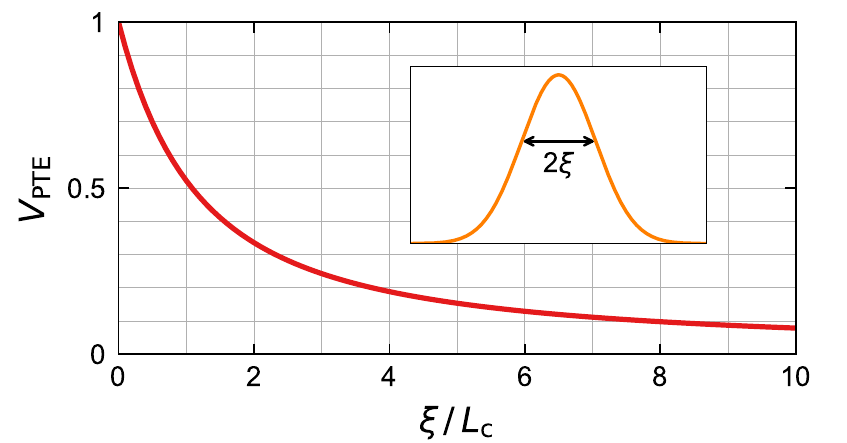}
\caption{Photovoltage as a function of the width of the incident light profile $\xi$, from Eq.\ \eqref{eq_vpte_gaussian}. The light profile width is normalized by the electron cooling length $\lc$, and $\vpte$ is normalized by its value at zero width. The inset shows the shape of the Gaussian optical profile.}
\label{fig_spotsize}
\end{figure}

To explore the upper limit of light confinement, and also for mathematical convenience, we start with an infinitely narrow light profile centered at $x=x_0$, $G(x) = \delta(x-x_0)$. The resulting electron temperature profile is
\begin{gather}
T_{\x{el,}\delta}(x;x_0) = \tc \exp\left(-|x-x_0|/\lc\right), \nonumber \\
\x{where} \label{eq_tel_delta} \\
\tc = \frac{\alpha \pin}{2\ka} \cdot \frac{\lc}{\wch}. \nonumber
\end{gather}
To account for a finite width of the optical mode in the waveguide, we now assume a Gaussian light profile,
\begin{equation}
G(x) = \frac{1}{\sqrt{2\pi}\xi} \exp\left(-x^2/2\xi^2\right),
\end{equation}
where $\xi$ defines the width of the light profile, as shown in the inset of Fig.\ \ref{fig_spotsize}. The electron temperature is then the convolution of this Gaussian light profile with the temperature arising from the delta-function light profile,
\begin{equation}
\tel(x) = \int {T_{\x{el,}\delta}(x;x_0) G(x_0)\x{d}x_0},
\label{eq_tel_broad}
\end{equation}
and the resulting photovoltage is
\begin{gather}
\vpte = 2S \cdot T_\xi, \nonumber \\
\x{where} \label{eq_vpte_gaussian} \\
T_\xi = \tc \, \x{erfcx}\left( \frac{\xi}{\sqrt{2}\lc} \right), \nonumber
\end{gather}
with $\x{erfcx}(x)$ the scaled complementary error function.

The scaling of the photovoltage with the light profile width is shown in Fig.\ \ref{fig_spotsize}. Here it can be seen that reducing the size of the light profile below the cooling length can significantly enhance the performance of PTE-based photodetectors. For the examples mentioned at the beginning of this section \cite{Schuler2016, Muench2019}, confining 1.55-$\upmu$m light to a spot width of $\sim$100 nm is responsible for a 2$\times$ to 3$\times$ enhancement of the photoresponse. For photodetectors operating in the mid-IR or beyond, this degree of light confinement could enhance the performance by 5$\times$ to 10$\times$.

\subsection{Channel length}

\begin{figure}[tbh]
\includegraphics{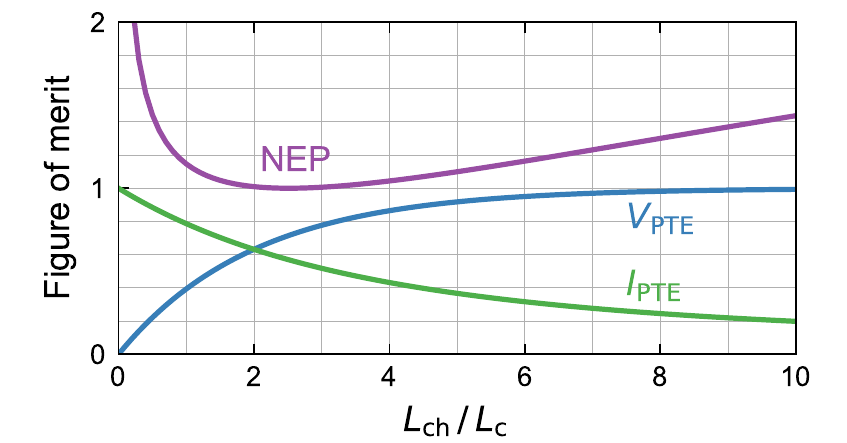}
\caption{Photodetector figures of merit as a function of the channel length $\lch$, from Eqs.\ \eqref{eq_vpte_finite}-\eqref{eq_nep_finite}. The channel length is normalized by the electron cooling length $\lc$, the photovoltage and photocurrent are normalized by their maximum values, and the NEP is normalized by its minimum value.}
\label{fig_lch}
\end{figure}

In the previous section we assumed that the distance between the source and drain contacts $\lch$ was much longer than the cooling length $\lc$, such that the integral in Eq.\ \eqref{eq_vpte} was taken from $-\infty$ to $+\infty$. Here we examine the impact of a finite channel length on photodetector performance. Assuming an ideal \textit{pn} junction as before, the photovoltage in Eq.\ \eqref{eq_vpte} becomes
\begin{equation}
\vpte = 2S \cdot \left[ \tel(0) - \tel(\lch/2) \right],
\end{equation}
where we have taken the light profile to be symmetric around $x=0$ such that $\tel(-\lch/2) = \tel(\lch/2)$. Following the approach of Eq.\ \eqref{eq_tel_broad} at $x = \lch/2$ gives
\begin{equation}
\vpte = 2S \cdot T_\xi \cdot \left( 1 - \e{-\lch/2\lc} \right).
\label{eq_vpte_finite}
\end{equation}
The resistance of the graphene channel is $R = (\lch/\wch) / \sigma$, where $\sigma$ is the conductivity, and from Eq.\ \eqref{eq_ipte} the photocurrent is thus given by
\begin{equation}
\ipte = 2S \cdot T_\xi \cdot \sigma \cdot \frac{\wch}{\lc} \cdot \left(\frac{1 - \e{-\lch/2\lc} }{\lch/\lc}\right).
\label{eq_ipte_finite}
\end{equation}
Similarly, from Eq.\ \eqref{eq_nep} the noise-equivalent power is
\begin{equation}
\x{NEP} = \frac{\pin}{2S \cdot T_\xi} \cdot \sqrt{\frac{\kb T}{\sigma} \frac{\lc}{\wch}} \cdot \left(\frac{\sqrt{\lch/\lc}}{1 - \e{-\lch/2\lc} }\right).
\label{eq_nep_finite}
\end{equation}
Each of these quantities is plotted in Fig.\ \ref{fig_lch} as a function of the channel length, relative to the cooling length. The photocurrent and photovoltage are normalized to their maximum values, while the NEP is normalized to its minimum value.

Figure \ref{fig_lch} makes it clear that the optimal channel length depends crucially on the sensing mode of the photodetector, with the photovoltage maximized at long $\lch$, the photocurrent maximized at short $\lch$, and the NEP minimized for an intermediate $\lch \approx 2.5\lc$. This latter value arises from the solution of the transcendental equation $1+x = \e{x/2}$, where $x = \lch/\lc$. This optimal value of the NEP arises from competition between the thermal noise and the photoresponse, as seen in Eqs.\ \eqref{eq_nep} and \eqref{eq_nep_finite}.

\subsection{Split-gate separation}

In the previous sections we assumed a perfect \textit{pn} junction at the point of light absorption, which is reasonable when the oxide thickness is larger than the spacing between the split gates. However, to achieve good electrostatic control over the graphene layer, thin oxide layers are ideal, and they are often thinner than the split-gate spacing. For example, in Ref.\ \citenum{Schuler2016} the split gates were embedded in the silicon waveguide with a spacing of 80 nm, while the oxide, composed of either Al$_2$O$_3$ or hBN, was $\sim$10 nm thick. Meanwhile, Ref.\ \citenum{Muench2019} utilized a traditional split-gate design, as shown in Fig.\ \ref{fig_pte}(d), with a split-gate separation of 100 nm and an oxide thickness of 40 nm.

In these situations, the split gates lose electrostatic control over the region between them, resulting in a uniform carrier density determined by the native doping of the graphene layer. Uniform doping means a uniform Seebeck coefficient, and thus via Eq.\ \eqref{eq_vpte} the region between the split gates will not contribute to the photoresponse when the temperature profile is symmetric around $x=0$. This may limit the overall performance of the photodetector and counteract the gains made by confining the light.

In this section we evaluate the photodetector performance when the oxide thickness is smaller than the split-gate separation, and compare it to the case of an ideal \textit{pn} junction. To model this situation, we approximate the Seebeck coefficient, thermal conductivity, and cooling length as piecewise functions of the form
\begin{equation}
\left\{ S(x),\ka(x),\lc(x) \right\} =
\left\{
\begin{matrix}
\left\{ S,\ka,\lc \right\} & , & x \le -\wg \\
\left\{ S_0,\ko,\lo \right\} & , & |x| < \wg \\
\left\{ -S,\ka,\lc \right\} & , & x \ge \wg
\end{matrix}
\right.
\label{eq_sx}
\end{equation}
where $\wg$ is half the spacing between the split gates. The values of $S$, $\ka$, and $\lc$ are determined by the voltage applied to the split gates, while the values of $S_0$, $\ko$, and $\lo$ are determined by the native doping of the graphene layer, which depends on the substrate, encapsulation, and device fabrication steps. The photovoltage from Eq.\ \eqref{eq_vpte} then becomes
\begin{align}
\vpte = S &\cdot \left[ \tel(\wg) + \tel(-\wg) \right] \nonumber \\
+ S_0 &\cdot \left[ \tel(\wg) - \tel(-\wg) \right].
\label{eq_vpte_full0}
\end{align}
Here we have assumed for simplicity that the channel length is long, as considering the impact of shorter channel length does not alter our main conclusions. In Eq.\ \eqref{eq_vpte_full0} we see that the photovoltage depends in general on both $S$ and $S_0$. However, the explicit dependence on $S_0$ disappears when the light spot is centered at $x=0$.

\begin{figure}[tbh]
\includegraphics{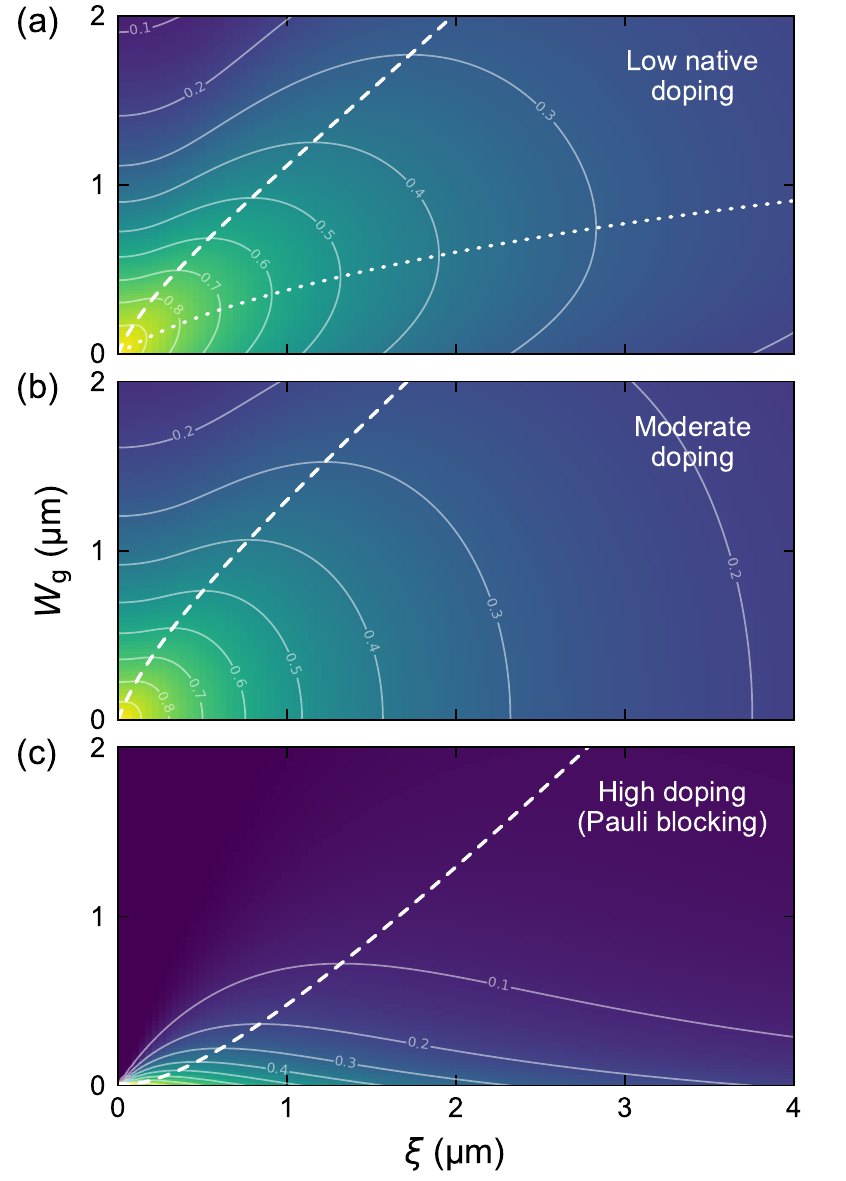}
\caption{Photovoltage $\vpte$ as a function of the light spot size $\xi$ and split-gate separation $\wg$, from Eq.\ \eqref{eq_vpte_full}. Panel (a) is for low native doping of the graphene layer, (b) is for moderate doping, and (c) is for high doping, when Pauli blocking has set in. The dashed lines show the value of $\xi$ at which $\vpte$ is maximized for each value of $\wg$, and vice versa for the dotted line in panel (a). The thin solid lines are contours of $\vpte$, relative to the maximal value at $(\xi=0,\wg=0)$.}
\label{fig_wg}
\end{figure}

To find the temperature at $x = \pm \wg$, we first assume an infinitely narrow light source incident at $x=x_0$. We then solve the heat equation in each region of the channel and match the temperature and the heat flux at $x = x_0$ and at $x = \pm\wg$. Finally, we convolve the temperature with a Gaussian light profile centered at $x=0$, giving
\begin{gather}
\vpte = 2S \cdot \tc \cdot \frac{\exp\left( \frac{\wg}{\lc} \right)}{\cosh\left(\frac{\wg}{\lo}\right)+r\sinh\left(\frac{\wg}{\lo}\right)} \nonumber \\
\cdot \left\{ \exp\left(-\frac{\wg}{\lc}\right) \cdot \exp\left(\frac{\xi^2}{2\lo^2}\right) \cdot \frac{1}{2}\left[\x{erf}(x_0^+)+\x{erf}(x_0^-)\right] \right. \nonumber \\
+ \left. \cosh\left(\frac{\wg}{\lo}\right) \cdot \exp\left(\frac{\xi^2}{2\lc^2}\right) \cdot \x{erfc}(x_c^+) \right\}, \nonumber \\
\x{where} \label{eq_vpte_full} \\
x_\text{0,c}^\pm = \left( \wg/\xi \pm \xi/L_\text{0,c} \right) / \sqrt{2}, \nonumber \\
r = \left( \lc/\lo \right) \cdot \left( \ko/\ka \right). \nonumber
\end{gather}
Given the complexity of this expression, there is not much insight to be gained through visual inspection. Thus, in Fig.\ \ref{fig_wg} we plot the photovoltage as a function of the spot size $\xi$ and the split-gate separation $\wg$.

In Fig.\ \ref{fig_wg} we consider three levels of native doping of the graphene layer between the split gates: low doping in panel (a), moderate doping in (b), and high doping in (c), where Pauli blocking has set in. In the region under the split gates, we assume the Fermi level has been tuned to $\ef = \pm 100$ meV, corresponding to a carrier density of $n \approx 10^{12}$ cm$^{-2}$. We also let $\lc = 1$ $\upmu$m, typical of experimental results \cite{Betz2013, Laitinen2014, Sierra2015}. Next, by assuming the electrical conductivity scales quadratically with $\ef$ (i.e., linearly with the carrier density) and the graphene density of states scales linearly with $\ef$, we find that $\ko/\ka = (\efo/\ef)^2$ and $\lo/\lc = \sqrt{\efo/\ef}$ \cite{Song2012}, where $\efo$ is the Fermi level in the region between the split gates.

In Fig.\ \ref{fig_wg}(a), we consider the case of low native doping by setting $\efo = 50$ meV, corresponding to $n_0 \approx 2.5 \times 10^{11}$ cm$^{-2}$. In panel (b) we consider moderate doping by setting $\efo = \ef =100$ meV, such that $\lo = \lc$ and $\ko = \ka$. Finally, in panel (c) we consider high native doping by setting $\efo = 300$ meV, corresponding to a carrier density $n_0 \approx 10^{13}$ cm$^{-2}$. We also investigate the role of Pauli blocking in this high-doping case, which is accomplished by eliminating the first term within the curly brackets in Eq.\ \eqref{eq_vpte_full}.

As expected, for all cases the photoresponse is maximized for the limiting case when $\xi = \wg = 0$. However, for finite values of $\xi$ or $\wg$, the photoresponse exhibits nonmonotonic behavior. This is indicated by the dashed line in each panel, which shows the value of $\xi$ that maximizes $\vpte$ for each fixed value of $\wg$. In contrast to Fig.\ \ref{fig_spotsize}, the photoresponse is not maximized at $\xi = 0$, but rather at $\xi \approx \wg$. Intuitively this makes sense: because the region between the split gates does not contribute to the photoresponse, reducing the light spot size much below $\wg$ can have a deleterious effect. In the other direction, making the light spot too large also reduces the photoresponse, following Fig.\ \ref{fig_spotsize}.

With respect to the optimal value of $\wg$, Figs.\ \ref{fig_wg}(b) and (c) indicate that a smaller $\wg$ is always better, as the region between the split gates does not contribute to the photoresponse. However, this trend does not hold for the low-doping case in panel (a). Here, the dotted line shows the value of $\wg$ that maximizes $\vpte$ for each fixed value of $\xi$. The fact that the optimal value of $\wg$ is finite is a consequence of the competing roles played by $\lo$ and $\ko$. Faster electron cooling between the split gates, corresponding to smaller $\lo$, reduces the electron temperature and thus the photoresponse. However, lower thermal conductivity slows the spread of heat away from the light source, which raises the electron temperature. This can be seen mathematically via the definition of $\tc$ in Eq.\ \eqref{eq_tel_delta}. Therefore, under the right conditions it is acceptable to have a finite $\wg$ and lower cooling length if these are counteracted by a small thermal conductivity.

Finally, let us briefly discuss the magnitude of the photoresponse in Fig.\ \ref{fig_wg}. In each panel, $\vpte$ is normalized to its maximal value at $(\xi=0,\wg=0)$, with the thin solid lines showing the contours of $\vpte$. Clearly, it is important to have $\wg,\xi \ll \lc$ for an optimal photoresponse. The photodetectors in Refs.\ \citenum{Schuler2016} and \citenum{Muench2019} have $\wg \approx \xi \approx 40$-$50$ nm, putting them at $\sim$$95\%$ of the optimal photoresponse in Figs.\ \ref{fig_wg}(a) and (b), but at only $\sim$$25\%$ in panel (c). Pauli blocking between the split gates should thus obviously be avoided -- this may not be an issue for photodetectors operating at 1.55 $\upmu$m, as native $p$-doping of graphene is typically in the range of 100-300 meV, but for the mid-IR and beyond this is a crucial design issue, even for cases where the light is strongly confined and the split-gate separation is small.

\subsection{Channel width}

Thus far we have examined photodetector performance considering transport along the length of the channel. In this section we examine in more detail the role of the channel width $\wch$, and determine its optimal value. To do so, we solve the 1D heat equation as before, but now for the temperature profile along the $y$-axis, perpendicular to the graphene channel. Averaging the temperature along the $x$-axis (over $\pm\lc$), the heat equation becomes
\begin{gather}
\frac{\x{d}^2\tel(y)}{\x{d}y^2} - \frac{\tel(y)}{\lc^2} + \frac{\beta \pin}{2\ka \lc} \frac{1}{\la} \e{-y/\la} = 0, \nonumber \\
\x{where} \label{eq_heat_y} \\
\beta = \int\limits_{-\lc}^{\lc}G(x)\x{d}x = \x{erf}\left(\frac{\lc}{\sqrt{2}\xi}\right) \nonumber
\end{gather}
when $G(x)$ is a Gaussian light profile.

The edges of the graphene channel play an important role in the resulting temperature profile. Recent experimental work has demonstrated strong energy relaxation at graphene edges, indicative of a resonant electron cooling mechanism \cite{Halbertal2017}. However, the edges in that study were highly disordered, while perfectly clean and straight edges may theoretically not exhibit any energy relaxation. To study these limiting cases, we solve Eq.\ \eqref{eq_heat_y} for two sets of boundary conditions: $\tel(y=0) = \tel(y=\wch) = 0$ for perfectly relaxing edges, and $\x{d}\tel/\x{d}y |_{y=0} = \x{d}\tel/\x{d}y |_{y=\wch} = 0$ for perfectly reflecting (i.e., nonrelaxing) edges.

The photovoltage is given by
\begin{equation}
\vpte = 2S \cdot \tav,
\label{eq_vpte_wch}
\end{equation}
where $\tav = \frac{1}{\wch}\int_{0}^{\wch}\tel(y)\x{d}y$ is the average temperature along the $y$-axis,
\begin{gather}
\tav \approx T_\xi \cdot \left. \left( \frac{\la}{\lc} - A \right) \middle/ \left( \frac{\la}{\lc} - \frac{\lc}{\la} \right) \right., \nonumber \\
\x{where} \label{eq_tav} \\
A =
\left\{
\begin{matrix}
\coth\left(\frac{\wch}{2\la} \right) \cdot \tanh\left(\frac{\wch}{2\lc} \right) & , & \x{relaxing edges} \\
\lc/\la & , & \x{reflecting edges}
\end{matrix}
\right. \nonumber
\end{gather}
and we have used $\x{erf}\left(\lc/\sqrt{2}\xi\right) \approx \x{erfcx}\left(\xi/\sqrt{2}\lc\right)$. Note that for reflecting graphene edges, the photovoltage reduces to what we obtained in Eq.\ \eqref{eq_vpte_gaussian}.

\begin{figure}[tbh]
\includegraphics{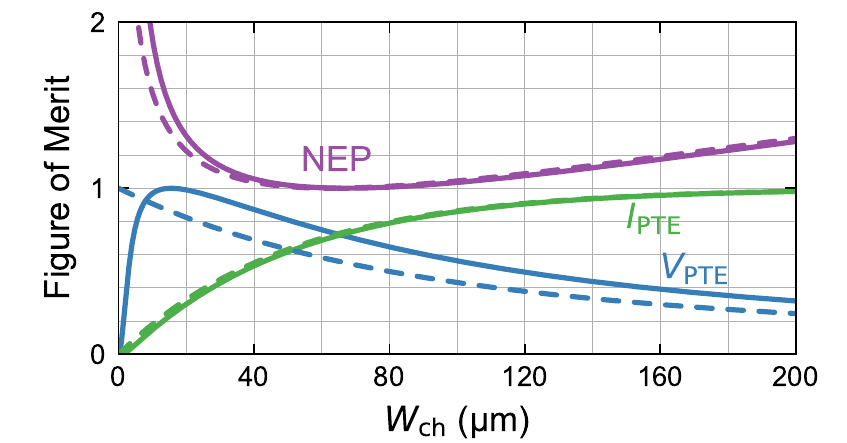}
\caption{Photodetector figures of merit as a function of the channel width $\wch$, from Eqs.\ \eqref{eq_vpte_wch}-\eqref{eq_nep_wch}. The photovoltage and photocurrent are normalized by their maximum values and the NEP is normalized by its minimum value. Here we have assumed a cooling length $\lc = 1$ $\upmu$m and an optical absorption length $\la = 50$ $\upmu$m. The solid (dashed) lines are for the case of perfectly relaxing (reflecting) graphene edges.}
\label{fig_wch}
\end{figure}

Using Eqs.\ \eqref{eq_ipte} and \eqref{eq_nep}, the photocurrent and the NEP are then given by
\begin{align}
\ipte &= 2S \cdot \tav \cdot \sigma \cdot \frac{\wch}{\lch}, \\
\x{NEP} &= \frac{\pin}{2S \cdot \tav} \cdot \sqrt{\frac{4\kb T}{\sigma}\frac{\lch}{\wch}}. \label{eq_nep_wch}
\end{align}
The photovoltage, photocurrent, and NEP are plotted as a function of the channel width in Fig.\ \ref{fig_wch}, with the solid (dashed) lines for the case of relaxing (reflecting) edges. Here we have assumed that $\lc = 1$ $\upmu$m and $\la = 50$ $\upmu$m. We see that the behavior of the graphene edges has little impact on the photocurrent or the NEP; the former is maximized for wide channels, as an increasing $\wch$ corresponds to a linear decrease in the channel resistance.

Meanwhile, the NEP is minimized for a channel width of $\wch \approx 65~\upmu\x{m}$, depending on the boundary conditions. For relaxing graphene edges, assuming $\wch,\la \gg \lc$, the optimal value of $\wch$ is given by the solution of $(1+2\zeta)(1+2x) = \e{x}$, where $\zeta = \lc/\la$ and $x = \wch/\la$. For the values in Fig.\ \ref{fig_wch}, this gives $\wch \approx 1.34\la \approx 67~\upmu\x{m}$. For reflecting edges, the optimal channel width is $\wch \approx 1.26\la \approx 63~\upmu\x{m}$, from the solution of $\e{x} = 1+2x$. Similar to the scaling of the NEP with channel length, this optimal value of the NEP arises from the nonmonotonic competition between the thermal noise and the photoresponse.

It is only for the photovoltage that the graphene edges play an appreciable role. For perfectly reflecting edges, $\vpte$ decreases monotonically with increasing $\wch$, a consequence of loss of PTE signal due to heat spreading along the $y$-axis. Meanwhile, for perfectly relaxing edges, $\vpte$ is maximized for an intermediate value of $\wch$, long enough to suppress the impact of heat relaxation at the edges but short enough to minimize loss of temperature due to heat spreading. Assuming $\wch,\la \gg \lc$, the optimal value of $\wch$ is given by the solution of $(1+2\zeta)(1+x) = \e{x}$. For the values in Fig.\ \ref{fig_wch}, this gives $\wch \approx 0.31\la \approx 15~\upmu\x{m}$.

\section{Material optimization}

So far we have demonstrated how to optimize the performance of PTE photodetectors in terms of the device geometry. Now we turn our focus to the properties of graphene itself, and how the material quality correlates with device performance. The photoresponse is typically quantified in terms of the photovoltage or photocurrent normalized by the input optical power,
\begin{gather}
R_V = \vpte / \pin, \\
R_I = \ipte / \pin.
\end{gather}

\subsection{Thermopower / thermal conductivity / \\ cooling length}

As shown in the previous sections, the material quantities that directly relate to photodetector performance are the Seebeck coefficient $S$, the electronic thermal conductivity $\ka$, and the electronic cooling length $\lc$. Each of these are related to the electrical conductivity, $\sigma$, through the relations \cite{Song2012, Sivan1986}
\begin{gather}
S = -\frac{1}{eT}\frac{\mathcal{K}_1}{\mathcal{K}_0}, \nonumber \\
\ka = \frac{2}{hT}\left( \mathcal{K}_2 - \frac{\mathcal{K}_1^2}{\mathcal{K}_0} \right), \nonumber \\
\lc = \sqrt{\ka / \gamma \cel}, \nonumber \\
\x{where} \label{eq_te_params} \\
\mathcal{K}_j = \int {\left( E-\ef \right)^j \widetilde{\sigma}(E) \left( -\frac{\partial f}{\partial E} \right) \x{d}E}, \nonumber
\end{gather}
$e$ is the electron charge, $h$ is Planck's constant, $\widetilde{\sigma} = \sigma/(2e^2/h)$, and $f$ is the Fermi-Dirac distribution.

The electron cooling is driven by a combination of supercollisions and direct electron-phonon relaxation, and is characterized by the cooling rate $\gamma$. Cooling arising from supercollisions and acoustic phonon emission is given by Eq.\ (13) of Ref.\ \citenum{Song2012}, while cooling due to optical phonon emission is given by Eq.\ (20) of the Supporting Information of Ref.\ \citenum{Pogna2021}.

The Seebeck coefficient is often approximated by the Mott formula, $S = (\pi \kb)^2T/(3\sigma) \cdot \x{d}\sigma/\x{d}E$, but we find that this expression severely overestimates the maximal value of $S$ for high-mobility graphene. Meanwhile, when $\ef > 2\kb T$ the thermal conductivity is fairly well approximated by the Wiedemann-Franz law, $\ka = \lwf \sigma T$, where $\lwf = (\pi \kb / e)^2/3$ is the Lorenz number. This relation generally holds for the situations we consider below. Note that for all calculations below we consider devices operating at room temperature, $T = 300$ K.

\subsection{Photovoltage}

Here we consider the impact that material quality has on the photovoltage. We are interested in the maximal photovoltage that can be obtained for a given material, so we consider a perfect \textit{pn} junction, a narrow light profile, and a long channel. We also compare relaxing and nonrelaxing graphene edges. For nonrelaxing edges the photovoltage is maximized for $\wch \rightarrow 0$, and from Eqs.\ \eqref{eq_tel_delta} and \eqref{eq_vpte_gaussian} the photoresponse is given by $R_V = S/\ka \cdot \lc/\la$. For relaxing edges, we use Eqs.\ \eqref{eq_vpte_wch} and \eqref{eq_tav}.

\begin{figure}[tbh]
\includegraphics{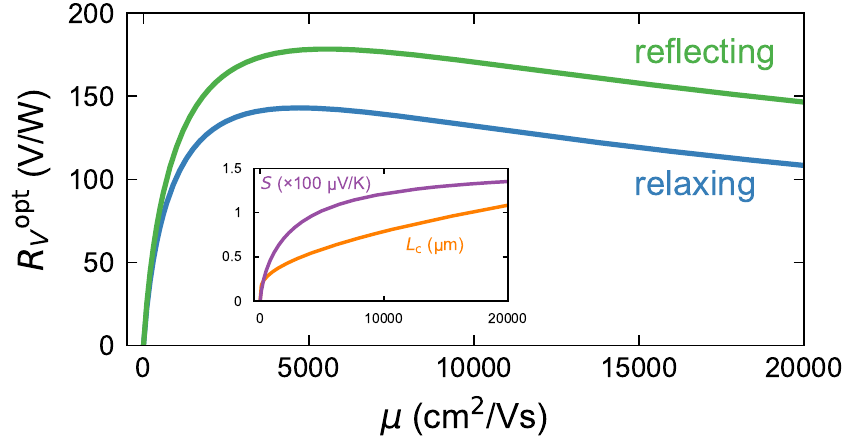}
\caption{Optimal photovoltage response as a function of graphene mobility, for the cases of relaxing and nonrelaxing edges. For nonrelaxing edges the optimal photovoltage is given by $\rvopt = \underset{\ef}{\max} \{S/\ka \cdot \lc/\la \}$. For relaxing edges, the optimal photovoltage is $\rvopt = \underset{\ef,\wch}{\max}\{2S \cdot \tav / \pin \}$, where $\tav$ is defined in Eq.\ \eqref{eq_tav}. The inset shows how the Seebeck coefficient and cooling length, at the point of optimal photoresponse, scale with graphene mobility.}
\label{fig_vpte_opt}
\end{figure}

In Fig.\ \ref{fig_vpte_opt}, we plot the maximal photoresponse as a function of graphene material quality, quantified by the carrier mobility $\mu$. For each case, we have written the conductivity as $\sigma(E) = \smin + \mu \cdot en(E)$, where $n(E) = E^2 / (\pi \hbar^2 \vf^2)$ is the carrier density and $\vf$ is the Fermi velocity. From this expression for $\sigma(E)$ we calculate $S$, $\ka$, and $\lc$ from Eq.\ \eqref{eq_te_params}, and find the maximal value of the photoresponse as a function of $\ef$ (and $\wch$ for the case of relaxing edges). We assume typical experimental values of $\smin = 0.2$ mS and $\la = 50$ $\upmu$m.

In Fig.\ \ref{fig_vpte_opt}, the optimal photoresponse reaches values of $100-200$ V/W, even for relatively low-mobility graphene. Meanwhile, to date the highest measured response of a waveguide-integrated photodetector in voltage mode is $R_V = 12.2$ V/W \cite{Muench2019}, indicating that there is a lot of room left for optimization of these devices. We discuss this device and another one in more detail in Section \ref{sec_exp}.

Figure \ref{fig_vpte_opt} is also striking in that the optimal photovoltage saturates with increasing graphene quality, reaching a maximum at $\mu \approx 5000$ cm$^2$/Vs. The reason for this is that the photoresponse is inversely proportional to the thermal conductivity $\ka$, which scales linearly with mobility. As discussed earlier, a higher thermal conductivity means faster diffusion of heat away from the \textit{pn} junction, reducing the electron temperature and thus the photoresponse. Meanwhile, as shown in the inset of Fig.\ \ref{fig_vpte_opt}, $S$ and $\lc$ scale sublinearly with $\mu$, yielding diminishing returns. In the case of $\lc$, at high mobility the electron cooling becomes limited by optical phonon emission \cite{Pogna2021}.

\subsection{Peltier cooling and the photocurrent}

For the photocurrent the picture is more complicated. Up to now we have only considered the generation of electrical signal via the Seebeck effect, but we have neglected other thermoelectric effects. Of particular relevance is the Peltier effect, which results in electronic cooling (or heating) when a current is driven across a junction between dissimilar materials. This is captured by adding an extra term to the heat equation, $\vec{j} \cdot \nabla \Pi$, where $\vec{j}$ is the electronic current density and $\Pi = S\cdot\left(\tel+T\right)$ is the Peltier coefficient \cite{Song2011}.

When the PTE photodetector is operated in an open-circuit configuration, the detected signal is the photovoltage, $\vpte$. In this configuration there is no current flow and thus the Peltier effect plays no role. However, in a closed-circuit configuration, a photocurrent $\ipte$ flows through the channel, leading to Peltier cooling at the \textit{pn} junction. The Peltier term in the heat equation then becomes $\vpte/(R\wch) \cdot T \cdot \x{d}S/\x{d}x$, where we assume low optical power such that $\tel \ll T$. Assuming a perfect \textit{pn} junction, and making use of Eqs.\ \eqref{eq_ipte} and \eqref{eq_vpte_ideal}, the 1D heat equation of Eq.\ \eqref{eq_heat_x} can be rewritten as
\begin{gather}
\frac{\x{d}^2\tel(x)}{\x{d}x^2} - \frac{\tel(x)}{\leff^2} + \frac{\alpha \pin}{\ka \wch} G(x) = 0, \nonumber \\
\x{where} \label{eq_heat_x_pelt} \\
\frac{1}{\leff^2} = \frac{1}{\lc^2} + \eta\frac{4 S^2 \sigma T}{\ka \lch}\delta(x), \nonumber
\end{gather}
and $\eta = 1-\exp(-\lch / 2\lc)$. Here we also assume reflecting graphene edges.

In this modified heat equation, $\leff$ thus describes an effective cooling length that is comprised of the usual cooling length, $\lc$, and an additional term arising from Peltier cooling at the \textit{pn} junction. Here we see that the Peltier cooling strength grows as $S^2$, arising from the Peltier term and the fact that $\ipte \propto S$. Thus, enhancing the Seebeck coefficient of graphene should yield diminishing returns and saturation of the photocurrent.

To quantify the role of Peltier cooling, we solve Eq.\ \eqref{eq_heat_x_pelt} by making use of Eq.\ \eqref{eq_vpte_full}. In Eq.\ \eqref{eq_vpte_full}, we had assumed a finite region in the middle of the graphene channel (of width $2\wg$) where the values of the thermal conductivity and cooling length $\left( \ko, \lo \right)$ differed from those under the split gates $\left( \ka, \lc \right)$. To capture the situation represented by Eq.\ \eqref{eq_heat_x_pelt}, we let $\ko = \ka$ and $\lo = \leff$. We write the Dirac delta function in the expression for $\leff$ as $\delta(x) = 1/(2\wg)$, and finally we take the limit $\wg \rightarrow 0$.

\begin{figure}[tbh]
\includegraphics{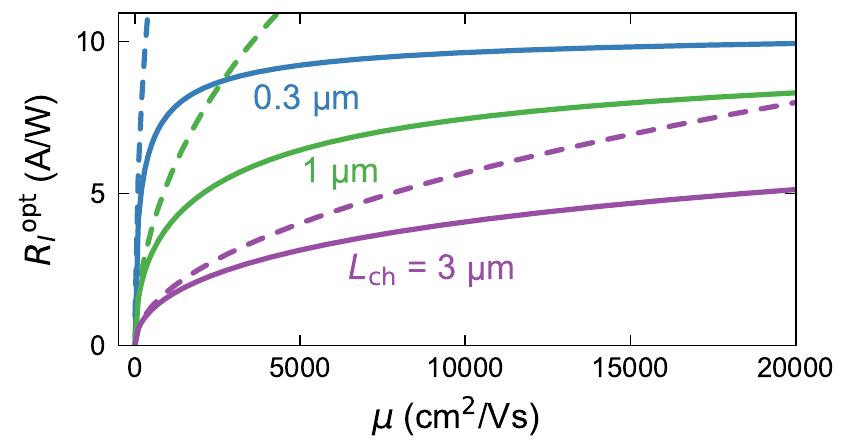}
\caption{Optimal photocurrent response as a function of graphene mobility, calculated as the maximum of Eq.\ \eqref{eq_ipte_peltier} over $\ef$. The solid (dashed) lines show the photoresponse with (without) Peltier cooling.}
\label{fig_ipte_opt}
\end{figure}

This yields a modified expression for the photocurrent response,
\begin{equation}
R_I = \left. S \cdot \eta \frac{\sigma}{\ka} \frac{\lc}{\lch} \middle/ \left( 1 + 2T \cdot \eta \frac{\sigma}{\ka}\frac{\lc}{\lch} \cdot S^2 \right) \right.,
\label{eq_ipte_peltier}
\end{equation}
where we have taken the limits $\xi \rightarrow 0$ and $\alpha \rightarrow 1$. The denominator captures the impact of Peltier cooling and indicates diminishing returns in the photoresponse with increasing graphene quality. This is shown in Fig.\ \ref{fig_ipte_opt}, where the solid lines indicate the optimal photocurrent response, calculated by maximizing Eq.\ \eqref{eq_ipte_peltier} over $\ef$, for three different channel lengths, $\lch = 0.3,1,3$ $\upmu$m. Similar to the photovoltage, the photocurrent response saturates with increasing graphene quality and does not show significant improvement for graphene mobilities above $\mu = 5000-10000$ cm$^2$/Vs. To highlight the impact of Peltier cooling, the dashed lines show the photoresponse when this effect is ignored. We attribute this strong Peltier effect to the unique electrical properties of graphene, which permit both high mobility and a high Seebeck coefficient.

\section{Comparison with Experiments}
\label{sec_exp}

Here we briefly compare the above models of photoresponse to a couple experimental examples of photodetectors, and discuss where room remains for performance optimization. We start with Ref.\ \citenum{Muench2019}, which used carefully designed split gates to confine the optical light profile to a few tens of nanometers. With this device design, the authors measured a photovoltage response of $R_V = 12.2$ V/W in one of their devices, which is the highest measured to date for waveguide-integrated photodetectors in voltage mode. Plugging their experimental parameters into the models presented above ($\smin = 0.2$ mS, $\mu = 2000$ cm$^2$/Vs, $\wg = \xi = 50$ nm, $\lch = 4.8$ $\upmu$m with 3.2 $\upmu$m covered by the split gates, $\wch = 1.7$ $\upmu$m), we would predict $R_V \approx 65$ V/W for nonrelaxing graphene edges and $R_V \approx 35$ V/W with relaxing edges. The former is more than $5\times$ higher than the measured value, while the latter is approaching the same order of magnitude (less than $3\times$). This suggests that thermal relaxation at the graphene edges may be limiting the already impressive device performance, due to the quite narrow channel width (also note that our models do not account for optical losses due to the metal split gates). Indeed, our models indicate that by increasing $\wch$ to $25$ $\upmu$m, the photoresponse can reach up to $56$ V/W.

Next we compare to Ref.\ \citenum{Miseikis2020}, which achieved high-frequency ($>$67 GHz) operation in a waveguide-integrated photodetector with high-mobility CVD graphene, and measured a photoresponse of $R_V = 6$ V/W. Using their device parameters ($\smin = 0.2$ mS, $\mu = 16000$ cm$^2$/Vs, $\wg = 500$ nm, $\xi = 300$ nm , $\lch = 10$ $\upmu$m with $8$ $\upmu$m covered by the split gates, $\wch = 80$ $\upmu$m), we predict $R_V = 32$ V/W, which is significantly larger than the measured value (more than $5\times$). The long channel width means that edge relaxation is not playing a role, so the discrepancy here may lie in the estimate of the cooling length $\lc$. As already commented by the authors of Ref.\ \citenum{Miseikis2020}, a cooling length of $\lc = 130$ nm was needed to fit their models to their measurements, and we find similar behavior with our models. This is quite a bit lower than the cooling length of $\lc \approx 1$ $\upmu$m predicted by the theory of supercollisions plus phonon emission \cite{Song2011, Pogna2021}. This discrepancy suggests that supercollisions and direct electron-phonon relaxation may not be enough to explain experiments, and one must consider the impact of resonant electron cooling \cite{Tikhonov2018, Kong2018, Halbertal2017}. Under this mechanism, a small number of resonant scatterers, arising from vacancy defects or covalent chemical adsorbates, may be sufficient to suppress the electronic heat while having an insignificant impact on the carrier mobility.

\section{Summary and Conclusions}
In summary, we provide a detailed analysis of the optimization of photodetectors based on the photothermoelectric effect in graphene. First, we show how various geometrical parameters -- including light spot size, channel length and width, and split-gate separation -- can tune the magnitude of the photocurrent, photovoltage, and noise-equivalent power.

Next, we demonstrate that the detector performance saturates with increasing graphene quality, and that this is true for both the photovoltage and the photocurrent. The photovoltage saturation is a consequence of the competing roles of the Seebeck coefficient and the cooling length on one side, and the thermal conductivity on the other. Meanwhile, the photocurrent saturates with increasing graphene quality because of Peltier cooling. This effect is often overlooked or ignored, but our analysis indicates it plays an important role in high-quality graphene, as a consequence of graphene's large mobility and Seebeck coefficient. In general, minimal improvements in optimal photoresponse are expected for mobilities above $5000$ cm$^2$/Vs.

Finally, comparisons of our models with experimental works suggest the important role played by resonant electron cooling in graphene. This mechanism can occur at graphene edges and can strongly limit photodetector performance in narrow channel devices. Such edge relaxation may be mitigated by proper choice of passivation, which may either delocalize resonant states responsible for the fast cooling, or move them to energies away from the regime of device operation. Resonant cooling may also be playing an important role in high-quality graphene, where electron cooling lengths are much shorter than what is predicted by the other cooling mechanisms. In our opinion, this issue deserves further detailed study.

\begin{acknowledgements}
\noindent We are grateful for useful discussions with Klaas-Jan Tielrooij and Max Lemme. ICN2 is supported by the Severo Ochoa Centres of Excellence program, funded by the Spanish State Research Agency (AEI, Grant No.\ SEV-2017-0706), and is funded by the CERCA program of the Generalitat de Catalunya. This work has received funding from the European Union's Horizon 2020 research and innovation program under Grant Agreement No.\ 825272 (ULISSES) and from the project “Waveguide-Integrated Mid-Infrared Graphene Detectors for Optical Gas Sensor Systems,” Reference No.\ PCI2018-093128, funded by the Spanish Ministry of Science and Innovation - AEI.
\end{acknowledgements}

\bibliography{optim_pte_graphene_bib_trunc}

\end{document}